\documentclass[a4paper,twocolumn,11pt,accepted=2021-03-29]{quantumarticle}

\pdfoutput=1

\usepackage[utf8]{inputenc}
\usepackage[english]{babel}
\usepackage[T1]{fontenc}
\usepackage{amsmath,amssymb,amsthm}
\usepackage{mathtools}
\usepackage{hyperref}
\usepackage{tikz}
\usepackage{multicol}
\usepackage[numbers,sort&compress]{natbib}

\usepackage{mathrsfs}
\usepackage{dsfont}
\usepackage{setspace}
\usepackage[export]{adjustbox}
\usepackage{makecell}

\usepackage{enumitem}
\usepackage{color}

\usepackage[ruled]{algorithm2e}

\SetCommentSty{mycommfont}

\newlength{\commentWidth}
\newcommand{\atcp}[1]{\tcp*[r]{\makebox[\commentWidth]{#1\hfill}}}
\newcommand{\btcp}[1]{\tcp*[f]{\makebox[\commentWidth]{#1\hfill}}}

\newtheorem{theorem}{Theorem}
\newtheorem*{theorem*}{Theorem}
\newtheorem{lemma}[theorem]{Lemma}

\usepackage{thmtools}
\usepackage{thm-restate}

\RequirePackage{filecontents}

\usepackage{mystyle}

\title{Combining hard and soft decoders for hypergraph product codes}
\author{Antoine Grospellier}
\affiliation{Inria, 2 Rue Simone IFF, CS 42112, 75589 Paris Cedex 12, France}
\author{Lucien Grou\`es}
\affiliation{Inria, 2 Rue Simone IFF, CS 42112, 75589 Paris Cedex 12, France}
\author{Anirudh Krishna}
\affiliation{Universit\'e de Sherbrooke, 2500 Boulevard de l'Universit\'e, Sherbrooke, QC J1K 2R1, Canada}
\author{Anthony Leverrier}
\affiliation{Inria, 2 Rue Simone IFF, CS 42112, 75589 Paris Cedex 12, France}

\begin{document}
\date{}

\twocolumn[
\begin{@twocolumnfalse}
    \maketitle
\begin{abstract}
    Hypergraph product codes are a class of constant-rate quantum low-density parity-check (LDPC) codes equipped with a linear-time decoder called small-set-flip ($\ssf$).
    This decoder displays sub-optimal performance in practice and requires very large error correcting codes to be effective.
    In this work, we present new hybrid decoders that combine the belief propagation ($\bp$) algorithm with the $\ssf$ decoder.
    We present the results of numerical simulations when codes are subject to independent bit-flip and phase-flip errors.
    We provide evidence that the threshold of these codes is roughly $7.5$\% assuming an ideal syndrome extraction, and remains close to 3$\%$ in the presence of syndrome noise. 
    This result subsumes and significantly improves upon an earlier work by Grospellier and Krishna ($\text{arXiv:}1810.03681$).
    The low-complexity high-performance of these heuristic decoders suggests that decoding should not be a substantial difficulty when moving from zero-rate surface codes to constant-rate LDPC codes and gives a further hint that such codes are well-worth investigating in the context of building large universal quantum computers.
\end{abstract}
\end{@twocolumnfalse}
]

\twocolumn
\section{Introduction}
It is imperative to make quantum circuits fault tolerant en route to building a scalable quantum computer.
The threshold theorem \cite{aharonov1997fault,kitaev1997quantum,knill1998resilient} guarantees that it will be possible to do so using quantum error correcting codes which encode information redundantly.
This redundancy serves as a buffer against errors but we need to be mindful of the trade-offs involved as the number of qubits we can control in the laboratory is limited.
A relevant  figure-of-merit to quantify this trade-off is the \emph{overhead}, defined as the ratio between the number of qubits in a fault-tolerant implementation  of a quantum circuit to the number of qubits in an ideal, noise-free environment.

Low-density parity-check (LDPC) codes are a natural class of codes to consider for implementations.
They are families of stabilizer codes $\C_n = \{\dsl n,k,d \dsr\}_n$ such that every stabilizer generator acts on a constant number of qubits and every qubit is involved in a constant number of generators \cite{gottesman1997stabilizer}.
Quantum architectures that would not satisfy these conditions would probably be very difficult to scale up, for instance because of difficulties to extract a syndrome fault-tolerantly.
A stronger restriction asks for quantum LDPC codes with geometric locality, where interactions only concern neighboring qubits in a 2 or 3-dimensional setup, but it is well known that this requirement severely restricts the ability of these codes to store information \cite{bravyi2010tradeoffs}.
On the other hand, general quantum LDPC codes can display a constant encoding rate $k/n = \Theta(1)$, while maintaining a large minimum distance $d = \Omega(\sqrt{n})$.
In a breakthrough paper, Gottesman exploited this favorable encoding rate and described a construction of fault-tolerant quantum circuits with constant space-overhead \cite{gottesman2014fault}.
This means that if we considered an ideal circuit that processes $m$ qubits, then its fault-tolerant counterpart will only require $\Theta(m)$ qubits.

Constructing good LDPC codes is difficult because we need to balance two competing constraints -- on the one hand, we want the weight of the stabilizers to be low, but on the other hand we want the stabilizers to commute.
Tillich and Z\'emor \cite{tillich2014quantum} proposed a construction called the hypergraph product code which overcomes this difficulty (see also generalizations \cite{kovalev2012improved,zeng2019higher}).
This construction takes two good \emph{classical} LDPC codes and constructs a quantum LDPC code with parameters $k = \Theta(n)$ and $d=\Theta(\sqrt{n})$.
In some sense, this construction generalizes the toric code and allows one to obtain a constant encoding rate while keeping the LDPC property as well as a large minimum distance.
As shown by Krishna and Poulin, there exists a framework rich enough to perform gates fault tolerantly on this class of codes \cite{krishna2019topological,krishna2019fault}.
Devising low-complexity decoding for hypergraph product codes is arguably one of the main challenges in the field right now.

Early analytic works establish and estimate the threshold of a broad class of LDPC codes with sublinear distance scaling \cite{kovalev2013fault, dumer2015thresholds} without the use of an efficient decoder.
In \cite{leverrier2015quantum}, Leverrier \textit{et al.} have shown the existence of a linear-time decoder for the hypergraph product codes called $\ssf$ and proved it corrects errors of size $O(\sqrt{n})$ in an adversarial setting.
In \cite{fawzi2018efficient}, Fawzi \textit{et al.} showed that the $\ssf$ decoder corrects with high probability a constant fraction of random errors in the case of ideal syndromes and later made these results fault tolerant by showing that this decoder is robust to syndrome noise as well \cite{fawzi2018constant}.
To be precise, they showed that $\ssf$ is a single-shot decoder and that in the presence of syndrome noise, the number of residual qubit errors on the state after decoding is proportional to the number of syndrome errors.
These works yield a rigorous proof of existence of a threshold for this class of codes, but only provide very pessimistic bounds on the numerical value of the threshold. 
Beginning with the seminal work of \cite{dennis2002topological}, statistical mechanical models have been used to make indirect estimates of the threshold of quantum error correcting codes \cite{bombin2012strong,chubb2018statistical}. Exploiting these ideas, Kovalev \textit{et al.}~\cite{kovalev2018numerical} showed that certain hypergraph product codes can achieve a relatively high threshold (approximately $7 \times 10^{-2}$) with the minimum-weight decoding algorithm.
Such an algorithm is too complex to be implemented in practice for general LDPC codes, however.

We note that some recent work from Panteleev and Kalachev investigated a quantum version of Ordered Statistical Decoding and obtained promising results for decoding small quantum LDPC codes \cite{bp_osd}.

\textbf{Related work:} 
Other families of quantum LDPC codes with constant rate include 2D and 4D hyperbolic codes: while the 2D version has a logarithmic minimum distance \cite{freedman2002z2, delfosse2013tradeoffs, breuckmann2016constructions}, the 4D hyperbolic codes satisfy $d = \Omega(n^c)$ for some $c>0$ and can therefore be interesting for fault-tolerance \cite{GuthLubotzky14, londe2019golden, breuckmann2020single, li2020numerical}.
Variants of these codes exhibit very good features and we compare our results to earlier works on hyperbolic codes.
The works \cite{conrad2018small, lavasani2019universal} study some properties of these codes further in the context of fault-tolerant quantum computation.

\textbf{Results and outline:}
In this paper, we present a new, efficient decoder for hypergraph product codes.
We combine the $\ssf$ decoder with a powerful decoder for classical LDPC codes called belief propogation ($\bp$).
The resulting decoders boast low decoding complexity, while at the same time yielding good peformance.
The idea behind these algorithms is to first decrease the size of the error using $\bp$, and then correct the residual error using the $\ssf$ decoder.
This paper subsumes and considerably improves upon an earlier work by Grospellier and Krishna \cite{grospellier2018numerical}.
We first study the performance of $\ssf$ by itself, and then study the performance of the hybrid decoder $\iter\bp + \ssf$.
Our simulations use a simple error model, that of independent bit-flip and phase-flip errors.
When compared to \cite{grospellier2018numerical}, the thresholds of codes are significantly improved (from $4.5$\% to roughly $7.5$\%).
We mention that there is not yet a theoretical threshold for the codes we focus on (obtained as a product of $(3,4)$-regular codes) together with the $\iter\bp + \ssf$ decoder.
Hence this is formally only a pseudo-threshold.
Furthermore, since the decoder is less demanding on the underlying quantum error correcting code, the weights of the stabilizers are reduced.
The stabilizers weights drop from $11$ to $7$.
We then extend this idea to decoding in the presence of syndrome noise.
In this model, we assume the syndrome bits are flipped with some probability in addition to qubits being subject to bit-flip and phase-flip noise.
We find that our codes perform well using a modified decoder called $\heur\bp + \ssf$.
The results are compared to the toric code, $2$D and $4$D hyperbolic codes.

In Section \ref{sec:background}, we begin by providing some background and establishing our notation.
We first review classical codes and discuss $\flip$ and the sum-product version of $\bp$ (simply referred to as $\bp$), and then proceed to review hypergraph product codes, and why naive generalizations of classical decoders $\flip$ and $\bp$ fail.
We introduce the $\ssf$ decoder and discuss how it overcomes these issues.
Section \ref{sec:idealsyndrome} then presents some results of numerical simulations.
Finally in Section \ref{sec:faultysyndrome} we discuss the results of simulations for faulty syndrome measurements.

\section{Background}
\label{sec:background}
\subsection{Classical codes}
\label{subsec:classicalcodes}
In this section, we shall review aspects of classical LDPC codes pertinent to quantum LDPC codes.
We begin by discussing the association between codes and graphs.
We then proceed to discuss expander graphs and the decoding algorithm $\flip$.
Finally we discuss a particular version of belief propagation ($\bp$) called the sum-product algorithm.

A classical code family $\{\C_n\}_n$, where $\C_n = \ker {H_n}$ is the binary linear code with parity-check matrix $H_n$, is said to be LDPC if the row weight and column weight of $H_n$ are upper-bounded respectively by constants $\Delta_C$ and $\Delta_V$ independent of $n$.
The weight of a row (or column) is the number of non-zero entries appearing in the row (or column).
In other words, the number of checks acting on any given bit and the number of bits in the support of any given check is a constant with respect to the block size.
These codes are equipped with iterative decoding algorithms (such as belief propagation) which have low time complexity and excellent performance.
Furthermore, they can be described in an intuitive manner using the factor graph associated with the classical code and for this reason these codes are also called graph codes.

The factor graph associated with $\C = \ker H$ is the bipartite graph $\G(\C) = (V \union C, E)$ where one set of nodes $V$ represents the bits (\textit{i.e.}, the columns of $H$) and the other set $C$ represents the checks (the rows of $H$).
For nodes $v_i \in V$ and $c_j \in C$, where $i \in [n]$ and $j \in [m]$, we draw an edge between $v_i$ and $c_j$ if the $i$-th variable node is in the support of the $j$-th check, or equivalently if $\h(i,j) = 1$.
It follows that a code $\C$ is LDPC if the associated factor graph has bounded degree, with left degree (associated with nodes in $V$) bounded by $\Delta_V$ and right degree bounded by $\Delta_C$.

Of particular interest are expander codes, codes whose factor graph corresponds to an expander graph.
Let $\G = (V \union C, E)$ be a bipartite factor graph such that $|V| = n$ and $|C| = m$ such that $n \geq m$.
We use $\Gamma(c)$ to denote the neighborhood of the node $c$ in the graph $\G$.
This naturally extends to a set $S$ of nodes; $\Gamma(S)$ includes any nodes connected to nodes in $S$ via an edge in $\G$.
Furthermore, $\deg(c) = |\Gamma(c)|$ is the degree of a node $c$.

The graph $\G$ is said to be $(\gamma_V,\delta_V)$-left-expanding if for $S \subseteq V$,
\begin{align}
    |S| \leq \gamma_V n \implies |\Gamma(S)| \geq (1-\delta_V)\Delta_V|S|~.
\end{align}
Similarly, the graph is $(\gamma_C,\delta_C)$-right-expanding if for $T \subseteq C$,
\begin{align}
    |T| \leq \gamma_C m \implies |\Gamma(T)| \geq (1-\delta_C)\Delta_C|T|~.
\end{align}
It is a \emph{bipartite} expander if it is both left and right expanding.

In their seminal paper, Sipser and Spielman \cite{sipser1994expander} studied expander codes and applied an elegant algorithm called $\flip$ to decode them.
They showed that if the factor graph is a left expander such that $\delta_V < 1/4$, then the $\flip$ algorithm is guaranteed to correct errors whose weight scales linearly with the block size of the code.
Furthermore, it does so in time scaling linearly with the size of the code block.

$\flip$ is a deceptively simple algorithm and it is remarkable that it works.
We describe it here as it forms the basis for the quantum case decoding algorithm $\ssf$.
Let $x \in \C$ be a codeword and $y$ be the corrupted word we receive upon transmitting $x$ through a noisy channel.
With each variable node $v_i$ in the factor graph, $i \in [n]$, we associate the value $y_i$.
With each check node $c_j$ in the factor graph, $j \in [m]$, we associate the syndrome bit $s_j = \sum_{i:v_i \in \Gamma(c_j)} y_i \pmod{2}$.
We shall say that a check node $c_j$ is unsatisfied if the syndrome is $1$ and satisfied otherwise.
Note that if $y \in \C$ is a codeword, then all the checks $c_j$, $j \in [m]$, must be satisfied.
Informally, $\flip$ searches for a variable node that is connected to more unsatisfied neighbors than it is to satisfied, and flips the corresponding bit.
This reduces the number of unsatisfied checks.
It is stated formally in Algorithm \ref{alg:flip} in Appendix \ref{app:flip and ssf} (comments in blue).

The algorithm can be shown to terminate in linear time.
For a detailed analysis, we point the interested reader to the original paper by Sipser and Spielman \cite{sipser1994expander}.

$\flip$ is not used in practice because it requires large code blocks to be effective \cite{richardson2008modern}; instead we resort to $\bp$.
In what follows, we shall use the sum-product algorithm and use $\bp$ to refer to this algorithm.
This algorithm is presented in Alg.~\ref{algo:BP}.

$\bp$ proceeds iteratively with $T$ iterations (described in Alg.~\ref{algo-SPstep}) further broken down into two elementary steps.
The first step (Alg.~\ref{algo:B2C}) involves variable nodes passing messages to checks and the second step (Alg.~\ref{algo:C2B}) exchanges the direction, and involves check nodes passing messages to variable nodes.
We introduce some notation to refer to these objects.
We let:
\begin{enumerate}
\item $p$ be the error probability on the variable nodes,
\item $s_j$ be the syndrome value of the check $c_j$ (0 if satisfied, 1 otherwise),
\item $m_{v_i \rightarrow c_j}^t$ be the message sent from variable-node $i$ to check-node $j$ on iteration $t$,
\item $m_{c_j\rightarrow v_i}^t$ be the message sent from check-node $j$ to variable-node $i$ on iteration $t$,
\item $\lambda^{t}_i$ is the approximate log-likelihood ratio computed at iteration $t$ for the variable-node $i$:
$\lambda^{t}_i > 0$ if it is more likely that the $i$-th variable node is more likely to be $0$ than $1$, otherwise $\lambda^{t}_i < 0$.
\end{enumerate}

On graphs with cycles, $\bp$ can only compute approximate values of the posterior probabilities.
However, it turns out to be relatively precise when the length of the smallest cycle (the girth) is big enough.
Thus the constraints of $\bp$ are weaker than that for $\flip$ and do not require expander graphs.

\subsection{Quantum codes}
\label{subsec:qtmcodes}
We now review the definition of the hypergraph product.
We proceed to discuss quantum expander codes, and the decoding algorithm proposed by Leverrier, Tillich and Z\'emor called $\ssf$.
We then present some earlier results of numerical simulations from \cite{grospellier2018numerical}.

CSS quantum codes are quantum error correcting codes that only contain stabilizers each of whose elements are all Pauli-$X$ operators (and identity) or all $Z$ \cite{calderbank1996good,steane1996multiple}.
The hypergraph product is a framework to construct CSS codes starting from two classical codes \cite{tillich2014quantum}.
The construction ensures that we have the appropriate commutation relations between the $X$ and $Z$ stabilizers without resorting to topology.
If the two classical codes are LDPC, then so is the resulting quantum code.
In general, the construction employs two potentially distinct bipartite graphs, but for simplicity, we shall only consider the product of a graph with itself here.
Let $\G$ be a bipartite graph, \textit{i.e.}, $\G = (V \union C, E)$.
We denote by $n := |V|$ and $m := |C|$ the size of the sets $V$ and $C$ respectively.

These graphs define two pairs of codes depending on which set defines the variable nodes and which set defines the check nodes.
The graph $\G$ defines the code $\C = [n,k,d]$ when nodes in $V$ are interpreted as variable nodes and nodes $C$ are represented as checks.
Note that $m \geq n - k$ as some of the checks could be redundant.
Similarly, these graphs serve to define codes $\tC = [m, \tk, \td]$ if $C$ represents variable nodes and $V$ the check nodes.
Equivalently, we can define these codes algebraically.
We say that the code $\C$ is the right-kernel of a parity check matrix $\h$ and the code $\tC$ is the right-kernel of the transpose matrix $\tH$.

We define a quantum code $\Q = \dsl n_{\Q},k_{\Q},d_{\Q}\dsr$ via the graph product of these two codes as follows.
The set of qubits is associated with the set $(V \times V) \union (C \times C)$.
The set of $Z$ stabilizers is associated with the set $(C \times V)$ and the $X$ stabilizers with the set $(V \times C)$. Ref.~\cite{tillich2014quantum} establishes the following:
\begin{lemma} The hypergraph product code $\Q$ has parameters:
\[ \llbracket n^2 + m^2, k^2 + (\tk)^2, \min(d, \td) \rrbracket.\]
\end{lemma}

Naively generalized to the quantum realm, both $\flip$ and $\bp$ perform poorly \cite{quant-pb}.
Unlike the classical setting, we are not looking for the exact error that occurred, but for any error belonging to the most likely error class since errors differing by an element of the stabilizer group are equivalent.
In the case of $\flip$, there exist constant size errors (typically half a generator) for which the algorithm gets stuck, which implies that $\flip$ will not work well even in a random error model.

\textbf{Overcoming the failure of $\flip$:} Leverrier \textit{et al.} \cite{leverrier2015quantum} devised an algorithm called small-set-flip ($\ssf$) obtained by modifying $\flip$.
This algorithm is guaranteed to work on quantum expander codes which are the hypergraph product of bipartite expanders.
The algorithm is sketched out in Alg.~\ref{alg:ssflip} in Appendix \ref{app:flip and ssf} (comments in blue).
For a detailed analysis of the algorithm, we point the reader to \cite{leverrier2015quantum}.
Note that this is not the full decoding algorithm -- it has to be run separately for both $X$ and $Z$ type errors.

Let $\F$ denote the union of the power sets of all the $Z$ generators in the code $\Q$.
For $E \in \field_2^{n^2 + m^2}$, let $\sigma_X(E)$ denote the \emph{syndrome} of $E$ with respect to the $X$ stabilizers.
The syndrome $\sigma_X(E) \in \field_2^{nm}$ is defined as $\h_X E$; the $j$-th element of this vector is $0$ if and only if the $j$-th $X$ stabilizer commutes with the error $E$.
Given the syndrome $\sigma_0$ of a $Z$ type error chain $E$, the algorithm proceeds iteratively.
In each iteration, it searches within the support of the $Z$ stabilizers for an error $F$ that reduces the syndrome weight. The case of $X$ errors follows in a similar way by swapping the role of $X$ and $Z$ stabilizer generators.

Ref.~\cite{leverrier2015quantum} proceeds to show that $\ssf$ is guaranteed to work if the graphs corresponding to classical codes are bipartite expanders.
They prove the following theorem (Theorem 2 in \cite{leverrier2015quantum}):
\begin{theorem}
    \label{thm:leverrierTillichZemor}
    Let $\G = (V \union C, E)$ be a $(\Delta_V, \Delta_C)$ biregular $(\gamma_V,\delta_V,\gamma_C,\delta_C)$ bipartite expander, with $\delta_V, \delta_C < 1/6$. Further suppose that $(\Delta_V, \Delta_C)$ are constants as $n$ and $m$ grow.
    The decoder \emph{$\ssf$} for the quantum code $\Q$ obtained via the hypergraph product of $\G$ with itself runs in time linear in the code length $n^2 + m^2$, and it decodes any error of weight less than
    \begin{align}
        w = \frac{1}{3(1+\Delta_C)}\min(\gamma_V n, \gamma_C m)~.
    \end{align}
\end{theorem}

\textbf{Overcoming the failure of $\bp$:} While $\bp$ can be adapted to decode quantum LDPC codes, it does not perform very well. The most common behaviour when $\bp$ fails at decoding quantum LDPC codes is that it does not converge: the likelihood ratios of some nodes keep oscillating.
This can be explained by the existence of some symmetric patterns in the Tanner graph which prevent $\bp$ from settling on a precise error.
To circumvent this, Poulin and Chung suggested some workarounds \cite{quant-pb} such as fixing the value of some qubits whose probabilities keep oscillating, or running $\bp$ again on a slightly modified Tanner graph where we randomly change the initial error probability of one of the qubits linked to an unsatisfied check.
The idea behind both of these solutions is to break the symmetry of the code.
While it does exhibit improvements, the results are still far from the performance of $\bp$ in the classical case.

An other approach to improve the performance of $\bp$ is to feed its output to a second decoder, with the hope that it will converge to a valid codeword if $\bp$ cannot.
This idea was recently investigated by Panteleev and Kalachev \cite{bp_osd} who considered a quantum version of the \emph{Ordered Statistical Decoding} algorithm $\texttt{OSD}$ \cite{fossorier1995soft}.
This algorithm was imported from the classical case where it is either used alone or after $\bp$.
The idea of this algorithm is to sort the different qubits by their log likelihood ratios, a measure of their reliability, before proceeding with a brute force approach.
Once $\texttt{OSD}$ has sorted the qubits, it will brute force \emph{all} valid corrections on the $w$ least reliable qubits, where $w$ is some tunable parameter, and then choose the most probable of these valid corrections (or fail if there are none).
If $w$ is proportional to the block-length, the time complexity is no longer polynomial.
Instead we can use the \texttt{OSD}-$0$ algorithm which is a simplified version that reduces the error floor of $\bp$.
In practice, this appears to work almost as well as \texttt{OSD}-$w$.
The time complexity is then polynomial, but may remain inappropriate for large codes.

In this work, we present some heuristic algorithms where the output of $\bp$ is fed to $\ssf$.
The idea behind these algorithms is to first decrease the size of the error using $\bp$ before correcting the residual error with the $\ssf$ decoder.
In practice, if $\bp$ manages to sufficiently decrease the error weight, then $\ssf$ will often reach a valid codeword without making a logical error.
We highlight that these hybrid algorithms have a time complexity far lower than that of $\texttt{OSD}$.

\section{Ideal syndrome extraction}
\label{sec:idealsyndrome}

\begin{figure*}[!t]
    \centering
    \includegraphics[width=\textwidth]{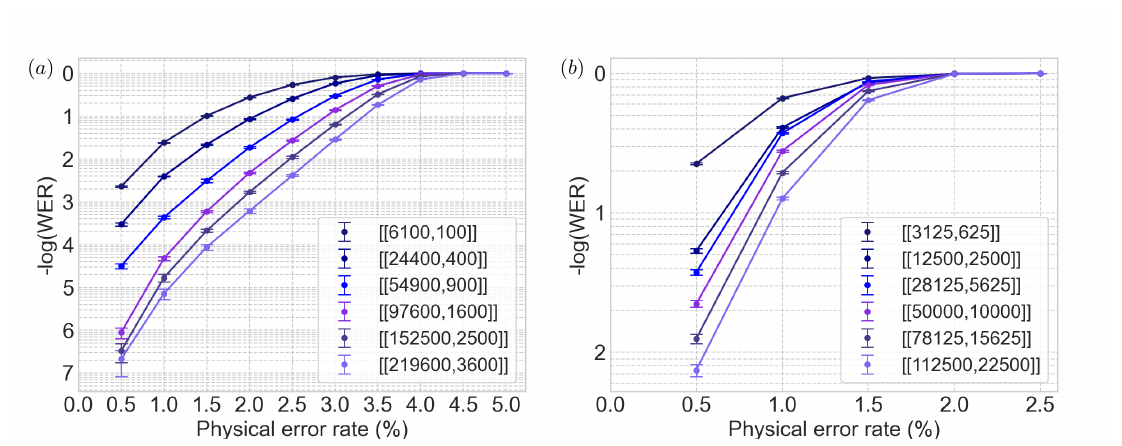}
    \captionof{figure}{
      Variation of word error rate (WER) with the physical error for hypergraph product codes formed as product of regular $(\Delta_V,\Delta_C)$-regular graphs with the $\ssf$ decoder from \cite{grospellier2018numerical}. Logarithms are base 10 throughout the paper.
    The errors bars indicate $99$\% confidence intervals, \textit{i.e.}, approximately $2.6$ standard deviations.
    $(a)$ Codes obtained as the product of $(5,6)$-regular graphs (encoding rate of $1/61 \approx 0.016$): we observe a threshold of roughly $4.5$\%.
    $(b)$ Codes obtained as the product of $(5,10)$-regular graphs (encoding rate of $0.2$): we observe a threshold of roughly $2$\%.
    }
    \label{fig:ssfonly}
\end{figure*}

In this section, we study a decoding algorithm called $\iter\bp + \ssf$.
To this end, we consider hypergraph product codes subject to a simple noise model.
We use classical codes generated with the configuration model, briefly described in Appendix \ref{app:configuration}.
We work with an independent bit-flip and phase-flip error noise model, where each qubit is afflicted independently by an $X$ or $Z$ error with probability $p$.
The advantage of studying such an error model with CSS codes is that it is sufficient to try to correct $X$ errors only to understand the performance of the whole decoding algorithm.
We focus here on ideal  syndrome measurements.
We will remove this assumption in the next section.

To establish a baseline, we begin by describing the performance of $\ssf$ as defined in \cite{leverrier2015quantum}.
Grospellier and Krishna \cite{grospellier2018numerical} studied the performance of $\ssf$ on quantum codes obtained as the hypergraph product of two $(5,6)$-regular graphs and $(5,10)$-regular graphs.

Fig.~\ref{fig:ssfonly} plots the logical error rate of these codes as a function of the physical error rate.
In this context, the logical error rate refers to the word error rate (WER), \textit{i.e.}, the probability that \emph{any} logical qubit fails.

In numerical benchmarks, we found a correlation between the performance of the classical codes under $\flip$ and the performance of the resulting quantum codes under $\ssf$.
The best among these codes were chosen as representatives for the quantum case and correspond to the different curves in the figure.
The $(5,6)$-regular codes have a threshold of roughly $4.5\%$, whereas the $(5,10)$-regular codes have a threshold of roughly $2$\%.
In this context, the threshold is the physical error rate below which we find that the logical error rate decreases as we increase the block size.
The error bars represent the $99$\% confidence intervals, \textit{i.e.}, approximately $2.6$ standard deviations.
At first glance, it appears that the $(5,10)$-regular codes perform much worse.
However this can be attributed to a much higher encoding rate compared to the first code family ($1/5$ versus $1/61$).

\begin{figure*}[!t]
   \centering
    \includegraphics[width=\textwidth]{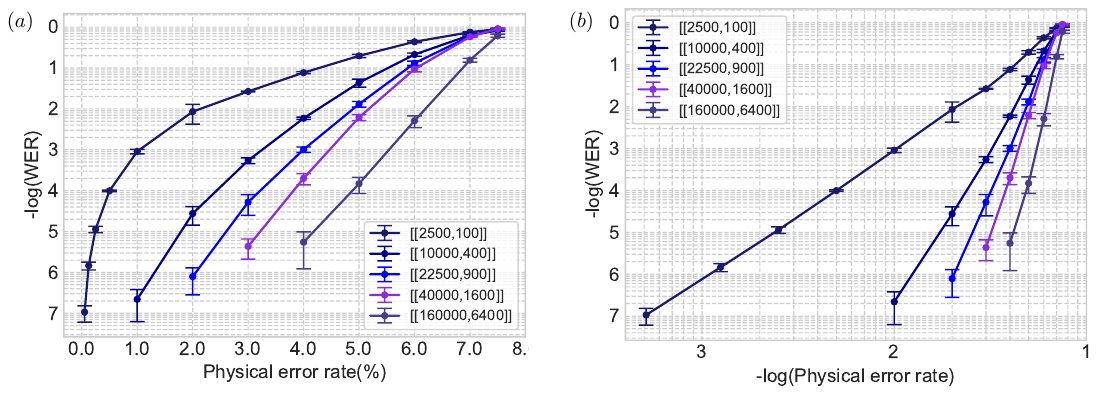}
      \caption{
        Variation of the word error rate (WER) with the physical error for a hypergraph product code formed as a product of $(3,4)$-regular graphs with the $\iter \bp+\ssf$ decoder. This code family displays an encoding rate of $0.04$.
        The error bars denote $99$\% confidence intervals, \textit{i.e.}, $\approx 2.6$ standard deviations.
        $(a)$ Base 10 logarithm of WER versus physical error rate.
        We see that the threshold is above 7\%.
        $(b)$ Log of WER versus log of physical error rate.
      }
      \label{fig:iterBPSSF}
\end{figure*}

Albeit promising, we note that $\ssf$ by itself requires large block sizes before it becomes effective.
This is unsurprising considering its classical counterpart also exhibits the same behaviour.
As mentioned in the previous section, this shortcoming of $\flip$ is addressed in the classical case by using instead soft decoding such as $\bp$. 
However, used naively, $\bp$ fails in the quantum realm. In practice, it fails at finding a valid codeword, but still manages to get rather close in the sense that the syndrome weight can be reduced by an order of magnitude.

Our idea to exploit this property is to start the decoding procedure with $\bp$ and switch to $\ssf$ after a certain number of rounds. For such an approach to work for a noisy syndrome, we need to specify a criterion to switch between the two decoders. 
Here, however, we consider a noiseless syndrome extraction and can apply the following simple iterative procedure: try decoding the error with $\ssf$ only; if this does not work, then perform a single round of $\bp$ followed by $\ssf$; if this still does not work, then perform 2 rounds of $\bp$ before switching to $\ssf$; and so on until a codeword is finally found, or when a maximum number $T_{\max}$ of $\bp$ rounds is reached.
In the latter case, we say that the decoder failed. 
This heuristic defines the hybrid decoder $\iter \bp+\ssf$ which is presented in Alg.~\ref{alg:iterbpssf} in Appendix \ref{app: variants of bp}.
To make a direct comparison with the $(5,6)$-regular codes presented above, we include the results of numerical simulations of the $(5,6)$-regular codes using the $\iter\bp+\ssf$ decoder in appendix \ref{app:56iterbpssf}. 

We now make some remarks concerning the time complexity of this algorithm. 
As described in Alg.~\ref{alg:iterbpssf} in Appendix \ref{app: variants of bp}, each iteration computes one  additional round of $\bp$.
To avoid redundancy in our computations, we save the output of the last step of each block of $\bp$ rounds.
When we proceed to the next iteration, we only need to compute the new round.
In practice most of the computation time is due to $\ssf$.
In our simulations, we choosed $T_{max} = 100$.

This hybrid decoder significantly improves upon the earlier results using only $\ssf$.
Fig.~\ref{fig:iterBPSSF} shows the variation of the WER versus the physical error rate using the hybrid decoder $\iter\bp+\ssf$.
The threshold appears to be at roughly $7.5$\%: below this value of physical error, the WER reduces as we increase the block size.
The log-log plot facilitates extrapolation to low noise rates.

\begin{table*}[t]
\centering
\begin{tabular}{|c|c|c|c|c|}
    \hline
     \textbf{Code} & \makecell{\textbf{Asymptotic}\\\textbf{Rate}} & \makecell{\textbf{Stabilizers}\\\textbf{weight} } & \textbf{Algorithm} & \textbf{Threshold} \\
    \hline
     Toric code \cite{wang2003confinement} & 0\% & 4 & \texttt{MWPM} & 10.5\% \\
     \hline
     \makecell{4,5-hyperbolic \\ code\cite{breuckmann2016constructions}} & 10\% & 4 and 5 & \texttt{MWPM} & 2.5\% \\
     \hline
        \makecell{$4D$-hyperbolic \\  code \cite{breuckmann2020single}} & 18\% & 12 & \bp & $\approx 5\%$ \\
     \hline
     (5,6) HGP code \cite{grospellier2018numerical} & 1.6\% & 11 & $\ssf$ & $\approx 4.6\%$ \\
     \hline
     {\color{blue}  (3,4) HGP code} &{\color{blue}  4\%} &{\color{blue}  7} & {\color{blue} $\iter\bp+\ssf$} &{\color{blue}  $\approx 7.5\%$ }\\
    \hline
\end{tabular}
\caption{
    Comparing different LDPC codes and decoders when subject to independent $X-Z$ noise and assuming ideal syndrome measurements.
    In this model, $p_x = p_z$.
    The results of this work highlighted in blue.}
\label{table:comparisons}
\end{table*}

Interestingly, the better performance of the hybrid $\iter\bp+\ssf$ decoder compared to the $\ssf$ decoder also comes with additional features such as an increased encoding rate (from $1.6\%$ to $4\%$) and a reduction of the stabilizer weight.
The $\ssf$ decoder used in \cite{grospellier2018numerical} indeed required classical codes generated from bipartite biregular factor graphs of degrees $(5,6)$.
The resulting quantum codes therefore had qubit degrees $10$ and $12$, and stabilizer weights $11$, respectively.
With the hybrid decoder, it suffices to use classical codes whose bipartite biregular factor graphs have degrees $(3,4)$.
The resulting quantum codes have qubit degrees $6$ and $8$, and stabilizer weights $7$, respectively.

This is surprising -- Theorem \ref{thm:leverrierTillichZemor} only guarantees performance of the $\ssf$ decoder if the graphs are sufficiently good expanders, which would require factor graphs with larger degrees that those we have considered.
Our hybrid decoder seems to be able to get away with a much lower expansion, and therefore smaller degrees.
This is important for physical implementations as higher degrees require more connectivity between different parts of the circuit.

Lastly, the word error rate is improved by several orders of magnitude, for a given block size.
Compare the codes $\dsl 24400, 400 \dsr$ generated from the $(5,6)$-regular family on Fig.~\ref{fig:ssfonly} and the $\dsl 22500, 900\dsr$ code generated from the $(3,4)$ family on Fig.~\ref{fig:iterBPSSF}. 
The encoding rate is twice as large in the second case and the code performance is significantly better for a given noise rate.
For instance at $p = 2$\%, the WER is $10^{-1}$ for the $\dsl 24400, 400 \dsr$ code but only  $10^{-3}$ for the $\dsl 22500, 900\dsr$ code.

Where do the $(3,4)$-regular codes with $\iter\bp+\ssf$ stand with respect to other codes?
In Table \ref{table:comparisons}, we compare their performance with the toric code, the $(4,5)$-hyperbolic code from \cite{breuckmann2016constructions}, the 4D hyperbolic code from \cite{breuckmann2020single}.
We find that the $(3,4)$-regular codes have a competitive threshold of roughly $7.5$\% only behind the toric code.
While the rate is not as good as that of \cite{breuckmann2020single}, it has a higher threshold and lower stabilizer weights.

\section{Dealing with syndrome noise}
\label{sec:faultysyndrome}

Although promising, the results of the previous section focus on an unrealistic problem since they assume perfect syndrome extraction.
We now move on to the more relevant setting where the syndrome themselves are error prone.
In addition to independent bit-flip and phase-flip noise each occurring at probability $p$, each of the syndrome bits is independently flipped with the same probability $p$. We choose the same probability for qubit and syndrome errors for simplicity.
Let us immediately note that we will not be able to use the $\iter\bp+\ssf$ decoder here since it requires knowledge of whether decoding has succeeded or not (\textit{i.e.}, whether the syndrome is null or not) in order to stop.
In the case where the syndrome is noisy, there is in general no way to know whether all qubit errors have been corrected.

Analyzing the performance of decoding algorithms with a noisy syndrome is not as straightforward as in the noiseless syndrome case, and we will in particular need to adapt our metrics.
We will follow the approach of Breuckmann and Terhal \cite{breuckmann2016constructions}.
When the syndrome is itself prone to error, we do not expect the output of the decoding algorithm to be an error-free code state.
We consider the following scenario corresponding to a quantum computation with $T$ layers of logical gates for instance\footnote{We could alternatively consider the problem of faithfully storing a state in a quantum memory for $T$ time steps for instance.}, and are interested in whether the final output is correct.
For each of these $T$ time steps, we consider both qubit noise (independent $X-Z$ noise with error rate $p$) and observe a noisy syndrome (corresponding to the ideal syndrome, with each bit further independently flipped with probability $p$).
After each time step, we use some efficient decoder $\Dec_1$  that returns some candidate error and apply the corresponding correction.
After the $T$ steps, we want to verify whether we are close to the correct codeword.
To this end, we perform error correction with the assumption that the syndrome can be noiselessly extracted.
This is because we are typically interested in a classical result and simply measure the qubits directly and compute the value of the syndrome directly (no need to measure ancilla qubits in the last round).
We then perform a final decoding procedure with a potentially different decoder $\Dec_2$.
We can then estimate the threshold as a function of $T$, and its asymptotic value corresponds to the so-called \emph{sustainable error rate} \cite{faultyperf}.
The idea is that if the physical error rate is below that threshold, then it means that one can perform arbitrarily long computations (or equivalently increase the lifetime of encoded information) by increasing the block length of the quantum codes.
Formally the sustainable error rate is defined as the failure rate obtained as a limit of the threshold error rate as the number of rounds of error correction is increased.
As mentioned above, the threshold itself is obtained as the block size of the code is increased to infinity such that the logical error rate is zero.

\begin{figure}[!h]
    \includegraphics[width=\columnwidth]{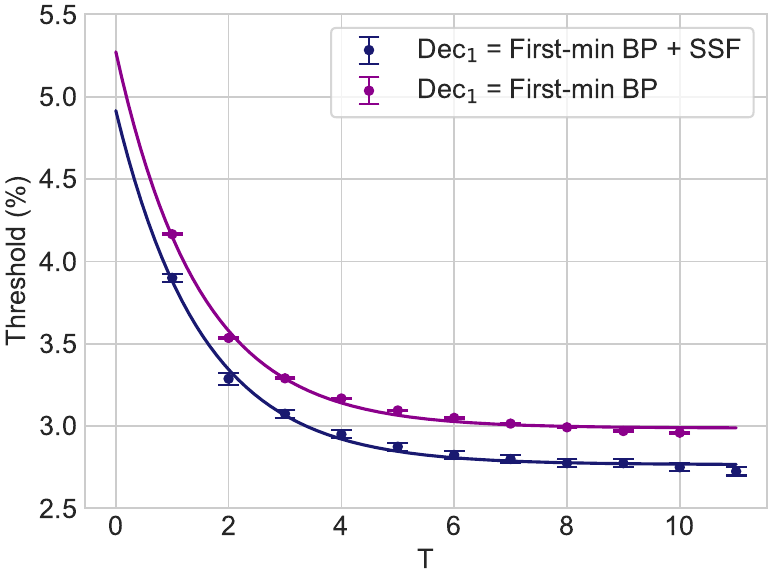}
    \caption{
        Evolution of the threshold as a function of $T$.
        Surprisingly, the simpler procedure $\heur \bp$ performs better than $\heur \bp +\ssf$ for $\Dec_1$ with a sustainable error rate which seems to be around 3\% (compared to 2.5\%).}
    \label{fig:noisythresholds}
\end{figure}

As alluded to above, $\iter\bp+\ssf$ is not a valid option for $\Dec_1$ since we would not know when to stop the decoding in general.
We have experimentally tried a number of heuristics for $\Dec_1$ and the one that performed the best is the $\heur \bp$ decoder (described in Alg.~\ref{alg:heurbp} in Appendix \ref{app: variants of bp}).
This decoder simply implements $\bp$ and terminates when the syndrome size stops decreasing.
To be precise, it performs a round of $\bp$, computes the estimated error and sees whether correcting for this error leads to a decrease of the syndrome weight.
If so, it continues with another round and otherwise, it returns the guess made at the previous round. 
We have numerically considered a number of variations for the stopping criterion but this one was consistently the best option. 
Another possibility that we investigated for $\Dec_1$ is to add $\ssf$ after $\heur \bp$.
This leads however to worse performance (as can be observed on Fig.~\ref{fig:noisythresholds}) and increases the decoder complexity. 

Having described the algorithm, we now discuss what parameters are fed to $\heur\bp$.
Recall from Alg.~\ref{algo:BP} that $\bp$ is initialized with prior information on how likely it is for each qubit to have been flipped.
This is done by specifying the Log-Likelihood Ratios (LLRs) $\{\lambda_i^0\}_{i=1}^{n}$ for every qubit.
If the syndromes are perfect, this algorithm proceeds to update these LLRs over several iterations.
To adapt $\bp$ to a fault-tolerant setting where it is employed $T$ times, we need to
\begin{enumerate}
    \item specify how to initialize the LLRs at each round, and
    \item how to process potentially incorrect syndrome information.
\end{enumerate}

In the fault-tolerant setting, qubits are subjected to i.i.d. noise only for the first round.
From the second round onwards, the noise is complicated and no longer Markovian;
in addition to the potential noise at a given round, the state of the qubits depends on the results of the corrections in all the previous rounds.
We make a simplifying assumption: when called at round $T$, $\bp$ is only fed approximate LLRs.
The LLRs are computed as if the qubits were subject to i.i.d. bit-flip and phase-flip errors with rate $p$, thereby ignoring all sources of noise from the previous rounds. 
For each qubit, these LLRs prescribe a bias $1-p$ to not being flipped, and $p$ to being flipped.

Secondly, although $\bp$ was described for perfect syndromes, it can easily be adapted to take the syndrome error into account.
We can simulate a Tanner graph with noisy checks by modifying the original Tanner graph.
For each check node in the graph, we add a unique variable node which represents whether or not it is erroneous.
We highlight some useful properties of this construction. 
These nodes do not create cycles, and therefore should not affect the performance of $\bp$.
They are treated like any other variable nodes, making this procedure easy to implement.
Since they are linked to only one node, and therefore always send the same message.
In particular, this adaptation of $\bp$ to the noisy syndrome case does not increase its time complexity.

For $\Dec_2$ on the other hand, the $\iter\bp+\ssf$ decoder would be admissible in principle since we assume that the syndrome is perfect at the last step.
For convenience, we chose to implement $\heur \bp+\ssf$ instead, but either options are expected to yield the same threshold.
A detailed review of the performance of $\heur \bp+\ssf$ is presented in Appendix \ref{app:heur}, and specifically in Fig.~\ref{fig:BPSSF}.

\begin{table*}[t]
\centering
\begin{tabular}{|c|c|c|c|c|c|}
     \hline
     \textbf{Code} & \makecell{\textbf{Asymptotic}\\\textbf{Rate}} & \makecell{\textbf{Stabilizers}\\\textbf{weight} } & \textbf{Algorithm} & \textbf{Threshold} & \makecell{\textbf{Single}\\\textbf{shot} } \\
    \hline
     Toric code \cite{wang2003confinement} & 0\% & 4 & MWPM & 2.9\% & No\\
     \hline
     \makecell{4,5-hyperbolic \\ code\cite{breuckmann2017hyperbolic}} & 10\% & 4 and 5 & MWPM & 1.3\% & No\\
     \hline
     {\color{blue} 3,4 HGP code} & {\color{blue} 4\%} & {\color{blue} 7} & {\color{blue} \makecell{$ \Dec_1: \heur \bp$\\ $\Dec_2: \heur \bp+\ssf$}} & {\color{blue} $\approx 3\%$ } & {\color{blue} Yes}\\
     \hline
\end{tabular}
\caption{
    Independent X-Z noise with syndrome errors $(p_x=p_z=p_{check})$.
    The results of this work highlighted in blue.
}
\label{table:comparisonsft}
\end{table*}

The noisy-sampling algorithm which gives the prescription to estimate the sustainable error rate is described in Alg.~\ref{alg:pseudoft} in Appendix \ref{app: noisy sampling}.
For brevity, we denote by $\mu(p)$ the Bernoulli distribution with bias $p$, \textit{i.e.}, drawing from this distribution $\Pr\{X = 1\} = p$ and $\Pr\{X = 0\} = 1-p$.

The simulations in the noisy syndrome case are far more time consuming than in the noiseless case since we need to plot the threshold as a function of $T$ to estimate its asymptotic limit.
To simplify the analysis, we again consider the independent $X-Z$ noise model: as before, we only need to simulate the case of $X$-errors only. 
This simplification does not affect the threshold.

Fig.~\ref{fig:noisythresholds} presents an estimate of the sustainable error rates for these algorithms. We obtain ``asymptotic" values around $2.5\%$ when $\Dec_1 =\heur \bp + \ssf$ and slightly above $3\%$ when $\Dec_1 = \heur \bp$.
The first algorithm seems to converge around $2.5$\% whereas the second one seems to converge above $3\%$.
A plausible explanation for the worse performance for the \textit{a priori} better algorithm $\heur \bp + \ssf$ is that $\bp$ does not manage to correct all syndrome errors, which is an issue for the $\ssf$ decoder.

As shown in Table~\ref{table:comparisonsft}, we obtain good results in comparison with other main families of quantum LDPC codes. 
Indeed the threshold is roughly the same as the toric code ($2.9$\% versus approx.~$3$\%).
While the stabilizer weights are lower, recall that the toric code has a zero asymptotic rate.
When compared to a positive rate code such as the $(4,5)$-hyperbolic codes, we cannot point to a clear winner: hyperbolic codes come with a better rate and lower stabilizer weights, but yield a lower threshold ($1.3\%$).

We conclude by pointing out that the sustainable error rate does not tell the whole story.
While not shown in Fig.~\ref{fig:noisythresholds}, the WER is typically very high in the vicinity of the threshold.
Similarly to the noiseless-syndrome case, very large codes are probably needed for the decoder to reach its full potential.

\section{Conclusion}

While previous studies showed that LDPC hypergraph product codes display good error suppression properties in the asymptotic regime, it is really the finite block length regime that matters for applications such as fault-tolerant quantum computation. We addressed this problem here by introducing new heuristic decoders for hypergraph product codes that combine soft information ($\bp$) and hard decisions ($\ssf$).
Our main motivation was that $\bp$ typically fails to converge when applied to quantum LDPC codes and that $\ssf$ typically only performs well when the syndrome has low weight.
Combining both decoders to let $\bp$ reduce the weight of the syndrome as much as possible, before turning to $\ssf$ to finish the decoding, leads to surprisingly good results. 
In the noiseless syndrome case, this combination yields an improvement from $4.6\%$ to $7.5\%$ for the threshold and much lower WER when compared to $\ssf$ alone, while at the same time relying on codes with higher encoding rate and lower stabilizer weights. 
In the noisy syndrome case, we studied a combination of $\bp$ and $\ssf$ where we use $\bp$ after each syndrome measurement to try to reduce the error and only rely on the hybrid decoder at the very last step of the procedure.
The sustainable error rate that we observe in simulations is competitive with the toric code as well as hyperbolic codes. 

LDPC codes are among the most versatile classical codes and come with efficient decoders with essentially optimal performance. For those codes, the hard decision decoder $\flip$ was successfully replaced by decoders such as $\bp$ that exploit soft information. It is tempting to believe that the same approach should also be true in the quantum case and that soft information decoders will convincingly replace decoders such as $\ssf$ in the future. While our results are a first step in this direction, they also call for a better understanding on how to process soft information in the case of quantum LDPC codes.

\section*{Acknowledgements}
We would like to thank Earl Campbell, Vivien Londe, David Poulin and Jean-Pierre Tillich for discussions.
AG, LG and AL acknowledge support from the ANR through the QuantERA project QCDA.
AK would like to thank the Fonds de Recherche du Québec Nature et Technologies (FRQNT) for the B2X scholarship.

\bibliographystyle{plainnat}
\bibliography{references}

\section{Appendix}
\renewcommand{\thesubsection}{\Alph{subsection}}

\subsection{\heur\bp + \ssf}
\label{app:heur}

In the noisy syndrome setting, we cannot expect $\ssf$ to know whether it has succeeded and therefore cannot apply $\iter\bp+\ssf$ anymore. Rather, we want to find a heuristic criterion to stop $\bp$ after the right number of rounds, and then feed the result to the $\ssf$ decoder. The simplest possibility is to observe the evolution of the syndrome weight through the successive rounds of $\bp$. This evolution is usually approximately periodic, displaying oscillations with the weight reaching a local minimum before increasing again.

We have empirically investigated several choices of stopping criterion, \textit{e.g.}, first minimum of the syndrome weight, global minimum in the 100 first rounds, and found that the best option was the first one. Stopping $\bp$ when the weight reaches its first minimum gives rise to our heuristic decoder $\heur \bp$. The $\heur \bp + \ssf$ decoder then corresponds to the case where the output of $\heur \bp$ is given to $\ssf$.
These algorithms are described in Alg.~\ref{alg:heurbp} and Alg.~\ref{alg:heurbpssf} in Appendix \ref{app: variants of bp}.

Fig.~\ref{fig:BPSSF} shows the variation of the WER versus the physical error rate for independent bit-flip and phase-flip noise, and ideal syndrome measurements. It is interesting to note that while the WER is degraded compared to $\iter\bp+\ssf$, the threshold behaviour is essentially identical for both decoding algorithms since they both yield a value around $7.5\%$. 
We had initially investigated thoroughly the $\heur \bp + \ssf$ decoder but the error floor occurring when the physical error rate approaches $1$\% led us to switch to $\iter\bp+\ssf$, for which this behaviour disappears. 
We found the good performance of $\heur \bp + \ssf$ remarkable given the simplicity of the heuristic, and it would be worthwhile studying other variants to decide reliably when to switch from $\bp$ to $\ssf$.

\begin{figure}[ht]
    \includegraphics[width=\columnwidth]{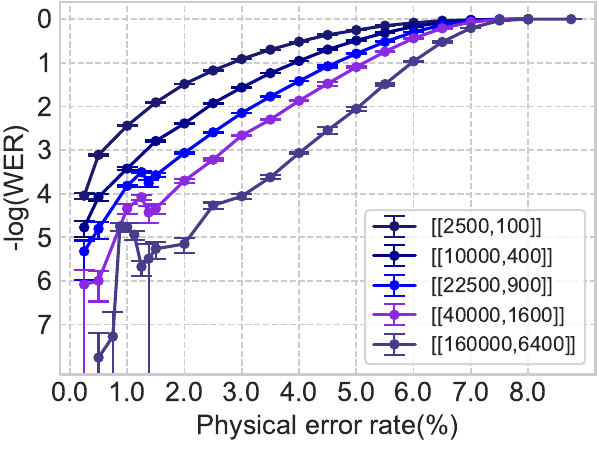}
    \caption{Word Error Rate (WER) as a function of the physical error for a hypergraph product code formed as a product of (3,4)-regular graphs with the $\heur \bp+\ssf$ decoder. The threshold is similar to that of $\iter\bp+\ssf$. The WER is worse, however, and odd patterns at low error rates.}
\label{fig:BPSSF}
\end{figure}

\subsection{Code construction}
\label{app:configuration}

We describe here how we construct the regular bipartite graphs we use in our simulations. Specifically we start with the \emph{configuration model} and then apply some post-processing to increase the girth of the factor graphs.
This algorithm generates graphs as follows:
\begin{enumerate}
  \item We first create an `empty' graph with the desired number of nodes.
        A vertex of degree $\Delta$ has $\Delta$ \emph{ports}.
        In our case, check nodes will have $\Delta_C$ ports and variable nodes will have $\Delta_V$ ports.
  \item We randomly draw edges between nodes by connecting a check port with a variable port.
        In the event we have two edges between a given pair of nodes, we randomly swap the double edges.
\end{enumerate}

This algorithm could potentially generate bipartite graphs with small cycles.
To avoid this, we use a post-processing algorithm that randomly swaps edges to increase the girth \cite{ccode-design}.
Although time consuming, this process yields good bipartite graphs.

\onecolumn

\subsection{\flip\ and $\ssf$}
\label{app:flip and ssf}


\setlength{\commentWidth}{12cm}

\begin{algorithm}[h!]
    \KwIn{Received word $y \in \field_2^n$}
    \KwOut{$w \in \field_2^n$, the deduced word}
  \vspace*{5pt}
  \hrule
  \vspace*{5pt}
     $w := y$      \atcp{Update $w$ iteratively}
     $\mathcal{F} = \emptyset$ \atcp{Flippable vertices}
    \For(\btcp{Setup phase: update bits}){$u \in V$}{
       If $u$ has more UNSAT than SAT neighbors, add it to $\mathcal{F}$\;
    }
    \While(\btcp{While flippable vertices exist}){$\exists u \in \mathcal{F}$}{
       flip $w_u$\;
       Decide whether the elements of $\Gamma(u)$ are UNSAT\;
       Decide whether the elements of $\Gamma(\Gamma(u))$ are in $\mathcal{F}$\;
    }
    \Return $w$\;
  \caption{\flip}
  \label{alg:flip}
\end{algorithm}

\setlength{\commentWidth}{13cm}

\begin{algorithm}[h!]
\DontPrintSemicolon
\caption{$\ssf$}
     \KwIn{A syndrome $\sigma_0 \in \field_2^{nm}$}
     \KwOut{Deduced error $\Ehat$ if algorithm converges and FAIL otherwise}
  \vspace*{5pt}
  \hrule
  \vspace*{5pt}
         $\Ehat = 0^{n_2 + m^2}$\atcp{Iteratively maintain $\Ehat$}
         $\sigma_0 = \sigma_X(E)$ \atcp{Iteratively maintain syndrome}
     \While{$\exists F \in \F : |\sigma_i| - |\sigma_i \oplus \sigma_X (F)| > 0$}{
       \bigskip
          $\displaystyle F_i = \arg \max_{F \in \F} \frac{|\sigma_i| - |\sigma_i \oplus \sigma_X(F)|}{|F|}$\\
          $\displaystyle \Ehat_{i+1} = \Ehat_i \oplus F$\\
          $\displaystyle \sigma_{i+1} = \sigma_i \oplus \sigma_X(F_i)$\\
       $\displaystyle i = i+1$
       \bigskip
        }
         \Return $\Ehat_i$ if $\sigma_X(\Ehat_i) + \sigma_0$ is zero and FAIL otherwise.
    \label{alg:ssflip}
\end{algorithm}


\newpage

\subsection{$\texttt{\bp}$ and subroutines}
\label{app: bp and subroutines}

\setlength{\commentWidth}{13cm}

\begin{algorithm}[h!]
\DontPrintSemicolon 
\caption{$\texttt{\bp}$} 
    \KwIn{%
        Time steps $T$\newline
        Syndromes $s \in \field_2^m$\newline
        Error probability $p$
        }
    \KwOut{Deduced error $\hat{v} \in \field_2^n$}
  \vspace*{2pt}
  \hrule
  \vspace*{5pt}
    \textbf{Initialization:} At time $t = 0$:
    $\displaystyle
    \forall v_i, \forall c_j \in \Gamma(v_i), m_{v_i\rightarrow c_j}^0= \lambda^{0}_i = \ln\left(\frac{1-p}{p}\right)~.
    $\;
    \For{$ 1\leq t \leq T$:}{
             $\texttt{sum-product-single-step}(t)$\;
        }
     \textbf{Terminate:} \atcp{Log-likelihood computation}
    \For{$i \in [n]$}{
             $\lambda^{t}_i =\lambda^{0}_i +\sum\limits_{c_{j}\in\Gamma(v_i)}m_{c_{j}\rightarrow v_i}^t$\;
            \eIf{$\lambda_i > 0$}{
                $\hat{v}_i = 0$\;
            }
            {
                $\hat{v}_i = 1$\;
            }
    }
    \Return $\hat{v}$
\label{algo:BP}
\end{algorithm}


\setlength{\commentWidth}{11cm}

\begin{algorithm}[h!]
\DontPrintSemicolon
\caption{$\texttt{sum-product-single-step}(t)$}
    \For(\btcp{Checks to bits}){$c_j \in C, v_i \in \Gamma(c_j)$}{
     \texttt{Check-to-bit}($c_j,v_i$)\;
    }
    \For(\btcp{Bits to checks}){$v_i \in V, c_j \in \Gamma(v_i)$}{
     \texttt{Bit-to-check}($v_i,c_j$)\;
    }
\label{algo-SPstep}
\end{algorithm}


\setlength{\commentWidth}{12cm}

\begin{algorithm}[h!]
\DontPrintSemicolon
    \caption{\texttt{Bit-to-Check}}
     \KwIn{Variable node $v_i$, check node $c_j \in \Gamma(v_i)$}
     \KwOut{$m_{v_i \rightarrow c_j}^{t+1}$}
  \vspace*{5pt}
  \hrule
  \vspace*{5pt}
     \Return
     $\displaystyle
     m_{v_i\rightarrow c_j}^{t+1}:=
     \ln\left(\frac{1-p}{p}\right)+\sum\limits_{c_{j'}\in\Gamma(v_i)\setminus{c_j}}m_{c_{j'}\rightarrow v_i}^t
     $
\label{algo:B2C}
\end{algorithm}

\setlength{\commentWidth}{12cm}

\begin{algorithm}[h!]
\DontPrintSemicolon
\caption{\texttt{Check-to-bit}}
     \KwIn{Check node $c_j$, variable node $v_i \in \Gamma(c_j)$}
     \KwOut{$m_{c_j \rightarrow v_i}^{t+1}$}
  \vspace*{5pt}
  \hrule
  \vspace*{5pt}
     \Return
     $\displaystyle m_{c_j\rightarrow v_i}^{t+1}:=
     (-1)^{s_j}2\tanh^{-1}\left(\prod\limits_{v_{i'}\in \Gamma(c_j)\setminus v_i}\tanh\left(\frac{m_{v_{i'}\rightarrow c_j}^t}{2}\right)\right)
     $
\label{algo:C2B}
\end{algorithm}

\subsection{Heuristics}
\label{app: variants of bp}

\setlength{\commentWidth}{12cm}

\begin{algorithm}[h!]
\DontPrintSemicolon
\caption{$\iter\bp+\ssf$}
\KwIn{%
  Syndrome $\sigma_0$ \newline
     maximal no. of $\bp$ iterations $T_{\max}$}
     \KwOut{ Deduced error $\Ehat$ if algorithm converges and FAIL otherwise.}
  \vspace*{5pt}
  \hrule
  \vspace*{5pt}
         $T=0$\;
     \While{$T\leq T_{\max}$}{
       \bigskip
             $\Ehat = \bp(T, \sigma_0, p)$\;
             $\sigma'=\sigma_0+\sigma_X(\Ehat)$\;
             $\Ehat=\Ehat + \ssf(\sigma')$\;
             $T = T+1$\;
             \Return $\Ehat$ if $\sigma_X(\Ehat) + \sigma_0$ is zero and keep running otherwise.\;
       \bigskip
        }
         \Return FAIL\;
    \label{alg:iterbpssf}
\end{algorithm}

\setlength{\commentWidth}{11cm}

\begin{algorithm}[h!]
\caption{$\heur\bp$}
\DontPrintSemicolon
\SetKwRepeat{Repeat}{do}{while}
     \KwIn{Syndrome $\sigma_0 = \sigma_X(E)$}
     \KwOut{Error $E'$ that minimizes the syndrome}
  \vspace*{5pt}
  \hrule
  \vspace*{5pt}
         $T=0$\;
        $\sigma_{\texttt{current}}=\sigma_0$\;
        $E_{\bp\texttt{current}}=0^n$\;
        \Repeat{$|\sigma_{\texttt{current}}| < |\sigma_{\texttt{prev}}|$\btcp{We repeat these steps while the syndrome decreases}}{
       \bigskip
             $T = T+1$\;
             $E_{\bp\texttt{prev}} = E_{\bp\texttt{current}}$\;
             $\sigma_{\texttt{prev}} = \sigma_{\texttt{current}}$\;
             $E_{\bp\texttt{current}} = \bp(T, \sigma_0, p)$\;
             $ \sigma_{\texttt{current}} = \sigma_0 + \sigma_X(E_{\bp\texttt{current}})$\;
       \bigskip
        }
         $E_{\bp}= E_{\bp\texttt{prev}}$\atcp{We choose the previous correction.}
         \Return $E_{\bp}$\;
    \label{alg:heurbp}
\end{algorithm}

\setlength{\commentWidth}{12cm}

\begin{algorithm}[h!]
    \caption{$\heur\bp + \ssf$}
    \DontPrintSemicolon
    \KwIn{Syndrome $\sigma_0 = \sigma_X(E)$}
    \KwOut{Deduced error $\Ehat$}
  \vspace*{5pt}
  \hrule
  \vspace*{5pt}
    $E_{\bp} = \heur\bp(\sigma_0)$\;
    $\sigma'=\sigma_0+\sigma_X(E_{\bp})$\;
    $E_{\ssf} = \ssf(\sigma')$\;
    $\Ehat = E_{\bp} + E_{\ssf}$\;
    \Return $\Ehat$\;
    \label{alg:heurbpssf}
\end{algorithm}

\subsection[]{Simulating decoding $(5,6)$-regular codes using $\iter\bp+\ssf$}
\label{app:56iterbpssf}

In this section, we present the result of simulating the hypergraph product codes obtained as a product of $(5,6)$-regular codes using the $\iter\bp+\ssf$ decoder.
As can be seen from the plot below, the threshold is improved, and is closer to $7$\% than the previously observed $5$\% as shown in fig. \ref{fig:ssfonly}.
On the other hand, these codes do not have as high a threshold as the $(3,4)$-regular codes which we presented above.

\begin{figure}[h]
\centering
\includegraphics[scale=0.7]{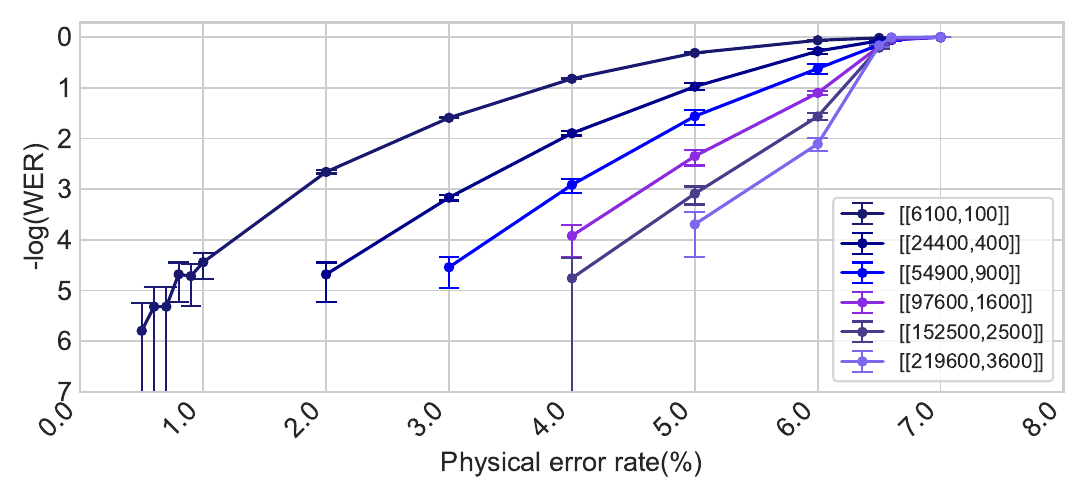}
\caption{Word Error Rate (WER) as a function of the physical error rate for a hypergraph product code formed as a product of $(5,6)$-regular graphs with the $\iter\bp + \ssf$ decoder.
The threshold is markedly better than that obtained from $\ssf$ alone.}
\label{fig:56iterbpssf}
\end{figure}

\setlength{\commentWidth}{13cm}

\subsection{Noisy sampling}
\label{app: noisy sampling}

\begin{algorithm}[h!]
\DontPrintSemicolon
 \caption{\texttt{Noisy-sampling}}
 \KwIn{%
   Bias probability $p$\newline
   Number of faulty rounds $T$}
     \KwOut{ SUCCESS if no logical errors, and FAIL otherwise}
  \vspace*{5pt}
  \hrule
  \vspace*{5pt}
     $E = 0^n$ \atcp{The code is initialized with no errors}
    \For{$i \in \{1,...,T\}$}{
       \bigskip
         Sample error $E_i$ from $\mu(p)^n$\;
          $E = E + E_i$\;
         Compute syndrome $\sigma_{i}=\sigma_{X}(E)$\;
         Sample syndrome noise $\widetilde{\sigma}_{i}$ from $\mu(p)^m$\;
         Let $\xi_{i} := \widetilde{\sigma}_{i}+ \sigma_{i}$\;
         Compute $\widehat{E}_i = \Dec_1(\xi_i)$\;
         $E = E + \widehat{E}_i$\;
       \bigskip
    }
  Sample error $E_{T+1}$ from $\mu(p)^n$ \;
  $E = E + E_{T+1} $\;
  Compute syndrome $\sigma_{T+1}=\sigma_{X}(E)$\;
  Compute $\Ehat_{T+1} = \Dec_2(\sigma_{X,T+1})$\;
  $E=\Ehat_{T+1} + E$\;
  \Return SUCCESS if $E$ is in the stabilizer group and FAIL otherwise\;
\label{alg:pseudoft}
\end{algorithm}

\end{document}